\documentclass[11pt]{article}
\usepackage{lscape}
\usepackage{rotating}
\usepackage{amsmath}

\begin{document}

\title{Relativistic Self-Consistent-Field Calculations of the Hyperfine
Structure in the 4d-Shell Ions}
\author{J.R. Persson \\
Department of Mathematics and Science, \\
University of Kristianstad, \\
S-291 88 Kristianstad,Sweden\\
jonas.persson@mna.hkr.se}
\maketitle

\begin{abstract}
Relativistic self-consistent-field calculations of the radial
hyperfine integrals have been performed in the 4d-shell element
ions. The comparison with available experimental results gives an
estimate of configuration interaction effects in the hyperfine
interaction in these ions. The results can also be used to derive
nuclear moments from laserspectroscopic measurements of radioactive
isotopes.

\end{abstract}


\section{Introduction}

During the last three decades a great deal of experimental\cite{1}
and theoretical\cite{2,3} work has been done on the hyperfine
structure (hfs) of the 4d-shell elements' atoms, whereas relatively
little is known about the hfs in the singly charged ions. In the
past years, the application of the laser-rf-double-resonance (LRDR)
technique\cite{4,5} as well as the use of saturation spectroscopy in
hollow-cathodes\cite{6}, has given experimental data of the hfs in
the Y and Zr ions. It is also expected that ion traps will open up a
possibility to perform measurements in these ions. Of particular
interest is it to perform systematic studies of the hfs in the
4d-shell atoms and ions since many-body effects are significant. In
the atoms many of these effects are masked by the problem of
obtaining accurate values of the eigenvectors used in the analysis.
The ions on the other hand are closer to Russel-Saunders (LS)
coupling, yielding a way of obtaining more accurate eigenvectors.
There is also an increased interest in spectroscopy of the 4d-shell
elements as a way of extracting nuclear properties, such as nuclear
moments and changes in the mean-squares charge radii, from
radioactive isotopes\cite{otten}. In most cases these studies are
performed using collinear laser spectroscopy on mass-separated
ion-beams or using ion-traps. As there are very little known of the
4d-shell ions, it is important to have as much data available as
possible. This becomes more important as some 4d-shell atoms only
have one stable isotope and in some cases with I=1/2, thus only
giving rise to magnetic dipole interaction. The best example is Y,
which is expected to exhibit drastic changes in the nuclear radii,
due to the shape transition at N=59. It is important to have high
quality calculations in order to extract the nuclear electric
quadrupole moment. In this article relativistic hyperfine integrals
of the magnetic dipole and electric quadrupole interactions for the
4d-shell ions, within the configurations $4d^{N}$, $4d^{N-1}5s$ and
$4d^{N-2}5s^{2}$, are presented. These integrals have been
calculated using relativistic (SCF) wave functions of Hartree-Fock
(HF) type and with a statistical exchange known as optimised
Hartree-Fock-Slater (OHFS)\cite{2}.

\section{Theoretical Approach}

Hyperfine structure analysis of the magnetic dipole interaction is normally
performed with the effective Hamiltonian taken as\cite{1,2}
\begin{equation}
H_{eff}^{1}=2{\frac{\mu _{0}}{{4\pi }}}\mu _{B}\sum_{i=1}^{N}[\mathbf{l}_{i}{%
\langle r^{-3}\rangle ^{01}}-\sqrt{1}0(\mathbf{sC}^{2})_{i}^{1}{\langle
r^{-3}\rangle ^{12}}+\mathbf{s}_{i}{\langle r^{-3}\rangle ^{10}}]\bullet
\mathbf{M}^{1}
\end{equation}%
where $\mathbf{M}^{1}$ is a nuclear tensor operator of rank 1, $\mathbf{{l}%
_{i}}$, ${(\mathbf{sC}^{2})_{i}^{1}}$ and $\mathbf{s}_{i}$ are the orbital,
spin-dipole and spin operators, respectively, of the open shell electrons,
with the summation over all open shells in the model space. In the case of
an unpaired s-electron only the spin operator term contributes to the hfs
energy. The effective Hamiltonian for the quadrupole interaction is usually
given as\cite{1,2}
\begin{equation}
H_{eff}^{2}={\frac{e}{4\pi \epsilon _{0}}}\sum_{i=1}^{N}[-\mathbf{C}_{i}^{2}{%
\langle r^{-3}\rangle ^{02}}+\sqrt{\frac{3}{10}}\mathbf{U}_{i}^{(11)2}{%
\langle r^{-3}\rangle ^{11}}+\sqrt{\frac{3}{10}}\mathbf{U}_{i}^{(13)2}{%
\langle r^{-3}\rangle ^{13}}]\bullet \mathbf{M}^{2}
\end{equation}%
where $\mathbf{M}^{2}$ is a nuclear tensor-operator of rank 2. $\mathbf{C}%
_{i}^{2}$ is a second-rank tensor-operator and ${\mathbf{U}^{(\kappa \lambda
)k}}$ are double tensor operators of rank $\kappa $ in spin space, rank $%
\lambda $ in orbital space and rank k in the combined spin-orbital space.
The effective radial integrals ${\langle r^{-3}\rangle _{nl}^{ij}}$ in (1)
and (2) are linear combinations of relativistic one-electron radial
integrals ${\langle r^{-3}\rangle _{nl}^{M}}$ , and ${\langle r^{-3}\rangle
_{nl}^{E}}$ [2]. In the non-relativistic limit the integrals ${\langle
r^{-3}\rangle _{nl}^{01}}$, ${\langle r^{-3}\rangle _{nl}^{12}}$ and ${%
\langle r^{-3}\rangle _{nl}^{02}}$ will approach the non-relativistic value
of $r^{-3}$, i.e.
\begin{equation}
\langle r^{-3}\rangle =\int P_{nl}^{2}(r)r^{-3}dr
\end{equation}%
For s-electrons the effective integral is defined by
\begin{equation}
\langle r^{-3}\rangle _{ns}^{10}={\frac{2}{3}}\langle r^{-3}\rangle _{ns}^{M}
\end{equation}%
which in the non-relativistic limit approaches the value
\begin{equation}
\langle r^{-3}\rangle _{ns}^{10}\rightarrow {\frac{2}{3}}\left[ {{\frac{%
dP_{ns}(r)}{{dr}}}}\right] _{r=0}^{2}
\end{equation}%
The ${\langle r^{-3}\rangle _{nl}^{10}}$, ${\langle r^{-3}\rangle _{nl}^{11}}
$ and ${\langle r^{-3}\rangle _{nl}^{13}}$ integrals have no
non-relativistic values and will approach zero. If we in addition to
relativistic effects also include configuration interaction effects, the
picture gets more complicated. However, it has been shown that the effective
Hamiltonians in (1) and (2) also act as effective operators in the case of
configuration interaction. The radial integrals should then be modified to
take this into account. This is normally done by adding a configuration
interaction contribution ($\Delta ^{ij}$) to the relativistic value.
\begin{equation}
{\langle r^{-3}\rangle _{nl,E}^{ij}}={\langle r^{-3}\rangle _{nl,R}^{ij}}%
(1+\Delta _{nl}^{ij})
\end{equation}%
The indices E and R refer to experimental and relativistic
Hartree-Fock values. In this way can configuration interaction
effects of one-body type be included in the effective Hamiltonian.
Configuration interaction effects of two-body type can be included
in the $\Delta ^{ij}$ corrections if they are allowed to be LS
dependent. A discussion of these two-body operators can be found in
\cite{7,8}. For the contact term ${\langle r^{-3}\rangle
_{nl,E}^{10}}$ and the pure relativistic terms ${\langle
r^{-3}\rangle _{nl,E}^{11}}$ and ${\langle r^{-3}\rangle
_{nl,E}^{13}}$, it is common to instead use.
\begin{eqnarray}
{\langle r^{-3}\rangle _{nl,E}^{10}} &=&{\langle r^{-3}\rangle _{nl,R}^{10}}+%
{\langle r^{-3}\rangle _{nl,C}^{10}} \\
{\langle r^{-3}\rangle _{nl,E}^{11}} &=&{\langle r^{-3}\rangle _{nl,R}^{11}}
\\
{\langle r^{-3}\rangle _{nl,E}^{13}} &=&{\langle r^{-3}\rangle _{nl,R}^{13}}
\end{eqnarray}%
Here the index C stands for core-polarisation. The reason to use (7)
instead of (6) for the contact term ${\langle r^{-3}\rangle
_{nl,E}^{10}} $ is that the dominant contribution is from
configuration interaction (core-polarisation) and not from
relativistic effects. The configuration interaction for the
${\langle r^{-3}\rangle _{nl,E}^{11}}$ and ${\langle r^{-3}\rangle
_{nl,E}^{13}}$ terms is usually assumed to be small compared with
the SCF-values and are therefore neglected. Normally one defines
effective radial parameters $a_{nl}^{ij}$ and $b_{nl}^{ij}$ which
are related to the nuclear moments $\mu _{I}$ and Q, respectively,
and to the effective values of the radial integrals ${\langle
r^{-3}\rangle }$ as:
\begin{eqnarray}
{a_{nl}^{ij}}{={\frac{2\mu _{B}}{{h}}}{\frac{\mu _{I}}{{I}}}{\langle
r^{-3}\rangle _{nl}^{ij}}} &&\text{{\ \ \ \ \ }}{l>0,ij=01,12,10} \\
{a_{ns}^{10}}{={\frac{2\mu _{B}}{{h}}}{\frac{\mu _{I}}{{I}}}{\langle
r^{-3}\rangle _{ns}^{10}}} && \\
{b_{nl}^{ij}}{={\frac{e^{2}}{{h}}}Q{\langle r^{-3}\rangle _{nl}^{ij}}} &&%
\text{{\ \ \ \ \ }}{l>0,ij=02,11,13}
\end{eqnarray}%
These parameters are treated as adjustable quantities to be fitted to the
experimental data in order to take configuration interaction and
relativistic effects into account. The experimentally determined
hfs-constants for a particular $|SLJ\rangle $ state are normally first
evaluated assuming J to be a good quantum number. Knowing these first order
values, it is possible to calculate the influence on the hfs from other J
states. This is normally done using perturbation theory to the second order,
why the corrections are refereed to as second order corrections (SOC). In
most cases these corrections are smaller than the experimental uncertainty
and can be neglected, but in the case of high-precision measurements or when
different J states are close energetically can the corrections be
substantial. An indication of large second order hyperfine interaction is
comparably large errors for the obtained experimental A and B constants. The
corrected hfs-constants can then be expressed as,
\begin{eqnarray}
{A(J)}{=\sum_{nl,ij=01,12,10}^{\mathit{all}\text{ }\mathit{config.}%
}k_{nl}^{ij}a_{nl}^{ij}} && \\
{B(J)}{=\sum_{nl,ij=02,11,13}^{\mathit{all}\text{ }\mathit{config.}%
}k_{nl}^{ij}b_{nl}^{ij}} &&
\end{eqnarray}%
in terms of (effective) radial parameters and an angular term ($k_{nl}^{ij}$
), that can be calculated using the eigenvectors of the states. In the 4d
shell ions a mixing exists between the configurations $4d^{N}$, $4d^{N-1}5s$
and $4d^{N-2}5s^{2}$. Due to this mixing will the magnetic dipole
interaction constants be expressed in twelve different radial parameters,
three from each configuration for the d-electrons, one from the unpaired
s-electron and two cross-configuration parameters. For the quadrupole
interaction eleven parameters are needed, three for the d-electrons from
each configuration and two cross-configuration terms. The
cross-configuration radial parameters are normally assumed to be small and
are therefore omitted from the analysis. For the determination of these
hyperfine parameters the hfs-constants should be known in a sufficient
number of atomic states in the three configurations. The parameters are
determined in a least-squares fit procedure. In most cases are the
hfs-constants only known in a few states. In order to reduce the number of
free parameters', assumptions have to be made to get experimental values of
the hyperfine parameters. For example, ratios between the spin-dipole and
orbital parameters, calculated with relativistic wavefunctions, can be used.
However, even if the hfs is known in ten or more states, the parameters
evaluated in the least-squares fit may be of rather unphysical magnitude.
This is especially the case for the parameters which coefficients, in the
parameterised expressions are small or sensitive to the intermediate
coupling constants, namely the spin-dipole and the relativistic quadrupole
operators $\mathbf{{U}^{(11)2}}$ and $\mathbf{{U}^{(13)2}}$ . The quadrupole
parameters $b_{nl}^{11}$ and $b_{nl}^{13}$ are in addition hard to extract
due to their small values. Another problem in the extraction of the
hyperfine parameters is that the expressions for the different hfs constants
are linear dependent in pure LS-coupling. It is only the breakdown of
LS-coupling that can resolve the linear dependence, which puts great demands
on the eigenvectors used. The calculations performed in this work was done
using relativistic wavefunctions of Hartree-Fock(HF) type, and wavefunctions
obtained by the so called Optimised Hartree-Fock-Slater (OHFS) method \cite%
{2}. Further description on the Self-Consistent-Field procedure, as well as
the methods used can be found in the review by Lindgren and Rosen \cite{2}.

\section{Results}

Using the relativistic wavefunctions calculated by the HF and OHFS SCF
procedures, hyperfine radial integrals have been obtained for the $4d^{N}$, $%
4d^{N-1}5s$ and $4d^{N-2}5s^{2}$ configurations in the 4d shell element
ions. The wavefunctions has been evaluated for the average energy of the
configurations. The results are presented in tables' 1-3. As a general trend
all the hyperfine integrals are increasing in magnitude with increasing
occupation number. The integrals ${\langle r^{-3}\rangle _{4d}^{01}}$ , ${%
\langle r^{-3}\rangle _{4d}^{12}}$ and ${\langle r^{-3}\rangle _{4d}^{02}}$
are found to increase when going from the configuration $4d^{N}$ to $%
4d^{N-2}5s^{2}$ for a certain element. The effective nuclear charge seen by
the d-electrons is increased due to less screening when an 4d-electron is
changed to an 5s-electron. The ${\langle r^{-3}\rangle _{4d}^{10}}$ , ${%
\langle r^{-3}\rangle _{4d}^{11}}$ and ${\langle r^{-3}\rangle _{4d}^{13}}$
integrals are more sensitive to relativistic effects and the contraction of
d-shells, and show a more irregular behaviour. From the calculated integrals
one finds an interesting feature, the ratio ${\langle r^{-3}\rangle
_{nd}^{11}}/{\langle r^{-3}\rangle _{nd}^{13}}$ seems to be fairly constant.
In addition, the change compared with similar calculations in the 3d to 5d
element atoms \cite{3,10} is very small, while the individual values change
drastically. The reason for this is not known Experimental data exists only
in Y and Zr, however, there might exist Fabry-Perot measurements but these
have not been considered here. The quality of the experimental hyperfine
integrals is depending on the number of states analysed and the quality of
the eigenvectors used.

\subsection{Y$^{+}$}

An extended analysis has been done in Y \cite{9} , where the odd
parity 4d5p configuration has also been analysed. The comparison
between the experimental and calculated integrals show that the HF
values describe the orbital operator better than the OHFS, following
the trend found in the 4d-shell atoms \cite{3}. It can also be seen
that the agreement for the orbital part is good, while the
spin-orbital integrals differ significantly. This result is by no
means surprising, as the spin-orbital part is more sensitive to the
quality of the eigenvectors as well as relativistic and
configuration interaction effects. The bad agreement for the
d-electron contact terms is due to core polarisation. However, the
problem with linear dependence as discussed earlier, is a very
severe problem, that also causes the contact terms to be linked to
each other. This problem can be resolved by obtaining more
experimental data. Yttrium is also a special case as it has only one
stable isotope with I=1/2, so one can not compare the hyperfine
interaction constants of radioactive isotopes with the stable
isotope in order to deduce the nuclear quadrupole moment. In this
case calculations of the radial hyperfine integrals, can be used to
deduce the nuclear quadrupole moment.

\subsection{Zr$^{+}$}

The hyperfine structure in Zr$^{+}$ has been studied experimentally
by Young et al.\cite{4}and theoretically by Beck and Datta
\cite{Beck}. The ionic ground state was measured by Campbell et
al.\cite{Campbell} \ The 12 states measured arises from the
$4d^{3}$and $4d^{2}5s$ configurations. In the first approximation is
this enought for an analysis of the 7 radial hfs parameters of these
configurations. However there are two complications, first the
states are mixed within the 3 configurations $4d^{3}$, $4d^{2}5s$ and $%
4d5s^{2}$, leading to 10 radial hfs parameters, secondly while some states
are heavily mixed, some are close to LS-coupling. This will lead to a linear
dependence of some parameters and that other parameters will have a very
small angular factor for some states. The equation-system to be solved will
be very badly conditioned, and the least-squares fit will be very sensitive
to small changes in the angular factors. This makes an analysis quite
uncertain, unless more states are measured. The experimental values of the
radial parameters presented in tables 2 and 3. are obtained from assuming
pure LS-coupling and excluding the most heavily mixed states. The result is
as expected quite bad.

\section{Conclusion}

The lack of experimental data makes it hard to draw any conclusions other
than the obvious. The 4d-shell elements are quite difficult to perform
experiments on, due to their refractoriness, and the short wavelength
transitions in the ions. There have, however, been recent developments on
different ion-sources as well as on lasers, so the available experimental
data is expected to increase within a not too distant future. One reason for
systematic studies is to find dependence of hfs effects depending on the
degree of ionisation. If the studies could be done on singly as well as
multiple charged ions, one could study the ${\langle r^{-3}\rangle _{4d}^{10}%
}$ integral that shows the extent of core-polarisation for non-s-electrons.
It has been shown in studies of ErI, ErII and ErIII that ${\langle
r^{-3}\rangle _{4f}^{10}}$ exhibits a near proportionality to the number of
electrons in the open shell \cite{11}. This is of importance as the
available ab initio methods fail to reproduce these effects.

\begin{landscape}
\begin{table}[tbph]
\caption{ Relativistic hyperfine integrals (in units of $a_{0}^{-3}$) for
the magnetic dipole and electric quadrupole interaction in the $%
4d^{N-2}5s^{2}$ configurations.}%
\begin{tabular}{cccccccccc}
&  &  &  & \multicolumn{3}{c}{Magnetic dipole} & \multicolumn{3}{c}{Electric
quadrupole} \\
Z & Ion & Conf. & Method & $\left\langle r^{-3}\right\rangle _{4d}^{01}$ & $%
\left\langle r^{-3}\right\rangle _{4d}^{12}$ & $\left\langle
r^{-3}\right\rangle _{4d}^{10}$ & $\left\langle r^{-3}\right\rangle
_{4d}^{02}$ & $\left\langle r^{-3}\right\rangle _{4d}^{13}$ & $\left\langle
r^{-3}\right\rangle _{4d}^{11}$ \\
40 & Zr & $4d5s^{2}$ & OHFS & 3.014 & 3.172 & -0.070 & 3.303 & 0.358 & -0.124
\\
&  &  & HF & 2.782 & 2.945 & -0.073 & 2.796 & 0.349 & -0.1127 \\
41 & Nb & $4d^{2}5s^{2}$ & OHFS & 3.804 & 4.012 & -0.091 & 3.828 & 0.478 &
-0.164 \\
&  &  & HF & 3.502 & 3.703 & -0.089 & 3.523 & 0.447 & -0.157 \\
42 & Mo & $4d^{3}5s^{2}$ & OHFS & 4.647 & 4.914 & -0.117 & 4.679 & 0.620 &
-0.210 \\
&  &  & HF & 4.309 & 4.584 & -0.123 & 4.336 & 0.604 & -0.215 \\
43 & Tc & $4d^{4}5s^{2}$ & OHFS & 5.551 & 5.887 & -0.147 & 5.592 & 0.785 &
-0.264 \\
&  &  & HF & 5.143 & 5.443 & -0.131 & 5.180 & 0.707 & -0.236 \\
44 & Ru & $4d^{5}5s^{2}$ & OHFS & 6.523 & 6.939 & -0.181 & 6.575 & 0.977 &
-0.328 \\
&  &  & HF & 6.073 & 6.462 & -0.171 & 6.120 & 0.904 & -0.305 \\
45 & Rh & $4d^{6}5s^{2}$ & OHFS & 7.570 & 8.078 & -0.221 & 7.634 & 1.200 &
-0.401 \\
&  &  & HF & 7.075 & 7.536 & -0.201 & 7.135 & 1.095 & -0.363 \\
46 & Pd & $4d^{7}5s^{2}$ & OHFS & 8.690 & 9.305 & -0.267 & 8.769 & 1.457 &
-0.485 \\
&  &  & HF & 8.145 & 8.687 & -0.235 & 8.219 & 1.314 & -0.428 \\
47 & Ag & $4d^{8}5s^{2}$ & OHFS & 9.891 & 10.627 & -0.320 & 9.987 & 1.752 &
-0.581 \\
&  &  & HF & 9.346 & 10.117 & -0.342 & 9.429 & 1.741 & -0.605%
\end{tabular}%
\label{t1}
\end{table}
\end{landscape}

\begin{landscape}
\begin{table}[tbph]
\caption{ Relativistic hyperfine integrals (in units of
$a_{0}^{-3}$) for the magnetic dipole and
electric quadrupole interaction in the $4d^{N-1}5s$ configurations.}%
\begin{tabular}{ccccccccccc}
&  &  &  & \multicolumn{4}{c}{Magnetic dipole} & \multicolumn{3}{c}{Electric
quadrupole} \\
Z & Ion & Conf. & Method & $\left\langle r^{-3}\right\rangle _{4d}^{01}$ & $%
\left\langle r^{-3}\right\rangle _{4d}^{12}$ & $\left\langle
r^{-3}\right\rangle _{4d}^{10}$ & $\left\langle r^{-3}\right\rangle
_{5s}^{10}$ & $\left\langle r^{-3}\right\rangle _{4d}^{02}$ & $\left\langle
r^{-3}\right\rangle _{4d}^{13}$ & $\left\langle r^{-3}\right\rangle
_{4d}^{11}$ \\
39 & Y & $4d5s$ & OHFS & 1.841 & 1.950 & -0.049 & 69.59 & 1.850 & 0.226 &
-0.084 \\
&  &  & HF & 1.779 & 1.863 & -0.037 & 60.86 & 1.788 & 0.190 & -0.065 \\
&  &  & Exp & 0.933 & 7.352 & 11.342 & 44.74 &  &  &  \\
40 & Zr & $4d^{2}5s$ & OHFS & 2.588 & 2.741 & -0.068 & 81.93 & 2.602 & 0.329
& -0.120 \\
&  &  & HF & 2.442 & 2.589 & -0.066 & 70.79 & 2.454 & 0.311 & -0.115 \\
&  &  & Exp & 2.74 & 0.95 & 7.20 & 37.44 &  &  &  \\
41 & Nb & $4d^{3}5s$ & OHFS & 3.359 & 3.564 & -0.091 & 93.68 & 3.379 & 0.448
& -0.160 \\
&  &  & HF & 3.137 & 3.320 & -0.081 & 82.118 & 3.156 & 0.404 & -0.143 \\
42 & Mo & $4d^{4}5s$ & OHFS & 4.179 & 4.444 & -0.117 & 105.06 & 4.206 & 0.587
& -0.207 \\
&  &  & HF & 3.910 & 4.169 & -0.116 & 92.21 & 3.934 & 0.559 & -0.202 \\
43 & Tc & $4d^{5}5s$ & OHFS & 5.056 & 3.390 & -0.148 & 116.48 & 5.091 & 0.749
& -0.262 \\
&  &  & HF & 4.722 & 5.018 & -0.131 & 103.05 & 4.755 & 0.675 & -0.232 \\
44 & Ru & $4d^{6}5s$ & OHFS & 5.998 & 6.412 & -0.183 & 127.90 & 6.043 & 0.938
& -0.325 \\
&  &  & HF & 5.620 & 6.003 & -0.169 & 113.54 & 5.662 & 0.865 & -0.300 \\
45 & Rh & $4d^{7}5s$ & OHFS & 7.012 & 7.520 & -0.224 & 139.33 & 7.069 & 1.158
& -0.399 \\
&  &  & HF & 6.584 & 7.072 & -0.217 & 124.18 & 6.365 & 1.093 & -0.382 \\
46 & Pd & $4d^{8}5s$ & OHFS & 8.102 & 8.717 & -0.270 & 150.87 & 8.173 & 1.411
& -0.484 \\
&  &  & HF & 7.626 & 8.138 & -0.222 & 134.70 & 7.695 & 1.234 & -0.404 \\
47 & Ag & $4d^{9}5s$ & OHFS & 9.268 & 10.006 & -0.342 & 162.93 & 9.355 &
1.702 & -0.580 \\
&  &  & HF & 8.787 & 9.554 & -0.324 & 147.79 & 8.862 & 1.688 & -0.500%
\end{tabular}%
\label{t2}
\end{table}
\end{landscape}

\begin{landscape}
\begin{table}[tbph]
\caption{ Relativistic hyperfine integrals (in units of
$a_{0}^{-3}$) for the magnetic dipole and
electric quadrupole interaction in the $4d^{N}$ configurations.}%
\begin{tabular}{ccccccccccc}
&  &  &  & \multicolumn{3}{c}{Magnetic dipole} &
\multicolumn{3}{c}{Electric
quadrupole} \\
Z & Ion & Conf. & Method & $\left\langle r^{-3}\right\rangle _{4d}^{01}$ & $%
\left\langle r^{-3}\right\rangle _{4d}^{12}$ & $\left\langle
r^{-3}\right\rangle _{4d}^{10}$ & $\left\langle r^{-3}\right\rangle
_{4d}^{02}$ & $\left\langle r^{-3}\right\rangle _{4d}^{13}$ & $\left\langle
r^{-3}\right\rangle _{4d}^{11}$ \\
39 & Y & $4d^{2}$ & OHFS & 1.506 & 1.605 & -0.045 & 1.512 & 0.198 & -0.077
\\
&  &  & HF & 1.492 & 1.569 & -0.034 & 1.499 & 0.167 & -0.060 \\
&  &  & Exp & 1.466 & 1.562 & -3.455 &  &  &  \\
40 & Zr & $4d^{3}$ & OHFS & 2.195 & 2.343 & -0.067 & 2.206 & 0.301 & -0.114
\\
&  &  & HF & 2.117 & 2.251 & -0.060 & 2.127 & 0.277 & -0.104 \\
&  &  & Exp & 2.09 & 1.27 & -4.38 &  &  &  \\
41 & Nb & $4d^{4}$ & OHFS & 2.927 & 3.129 & -0.091 & 2.943 & 0.419 & -0.157
\\
&  &  & HF & 2.784 & 2.954 & -0.076 & 2.799 & 0.369 & -0.133 \\
42 & Mo & $4d^{5}$ & OHFS & 3.711 & 3.975 & -0.118 & 3.733 & 0.557 & -0.205
\\
&  &  & HF & 3.521 & 3.769 & -0.112 & 3.542 & 0.521 & -0.193 \\
43 & Tc & $4d^{6}$ & OHFS & 4.556 & 4.892 & -0.150 & 4.585 & 0.718 & -0.262
\\
&  &  & HF & 4.306 & 4.602 & -0.132 & 4.334 & 0.647 & -0.231 \\
44 & Ru & $4d^{7}$ & OHFS & 5.464 & 5.883 & -0.187 & 5.503 & 0.906 & -0.327
\\
&  &  & HF & 5.170 & 5.554 & -0.172 & 5.207 & 0.835 & -0.300 \\
45 & Rh & $4d^{8}$ & OHFS & 6.443 & 6.958 & -0.229 & 6.492 & 1.124 & -0.403
\\
&  &  & HF & 6.096 & 6.615 & -0.234 & 6.140 & 1.095 & -0.404 \\
46 & Pd & $4d^{9}$ & OHFS & 7.497 & 8.122 & -0.278 & 7.559 & 1.375 & -0.489
\\
&  &  & HF & 7.109 & 7.605 & -0.217 & 7.172 & 1.172 & -0.391%
\end{tabular}%
\label{t3}
\end{table}
\end{landscape}

\end{document}